# Mobile Agent as an Approach to Improve QoS in Vehicular Ad Hoc Network

Rakesh Kumar,
Assistant Professor, M. M. University, Mullana
Haryana, India,

Dr. Mayank Dave
Associate Professor, N.I.T. Kurukshetra
Haryana, India

## ABSTRACT
Vehicular traffic is a foremost problem in modern cities. Huge amount of time and resources are wasted while traveling due to traffic congestion. With the introduction of sophisticated traffic management systems, such as those incorporating dynamic traffic assignments, more stringent demands are being placed upon the available real time traffic data. In this paper we have proposed mobile agent as a mechanism to handle the traffic problem on road. Mobile software agents can be used to provide the better QoS (Quality of Service) in vehicular ad hoc network to improve the safety application and driver comfort.

## 1. INTRODUCTION
A Vehicular Ad hoc Network (VANET) is a form of wireless ad hoc network to provide communications among vehicles and nearby roadside equipments. It is emerging as a new technology to integrate the capabilities of new generation wireless networking to vehicles. The major purpose of VANET is to provide (1) ubiquitous connectivity while on the road to mobile users, who are otherwise connected to the outside world through other networks at home or at the work place, and (2) efficient vehicle-to-vehicle communications that enable the Intelligent Transportation Systems (ITS). ITS includes a variety of applications such as cooperative traffic monitoring, control of traffic flows, blind crossing (a crossing without light control), prevention of collisions, nearby information services, and real-time detour routes computation. In this paper nodes and vehicles are used interchangeably.

## 2. NETWORK ARCHITECTURES AND CHARACTERISTICS
Wireless ad hoc networks generally do not rely on fixed infrastructure for communication and dissemination of information. VANETs follow the same principle and apply it to the highly dynamic environment of surface transportation.

As shown in Fig. 1, the architecture of VANETs mainly falls within three categories: pure cellular/WLAN, pure ad hoc, and hybrid.

VANETs may use fixed cellular gateways and WLAN/WiMax access points at traffic intersections to connect to the Internet, gather traffic information, or for routing purposes. The network architecture under this scenario is a pure cellular or WLAN structure as shown in Fig. 1(a). VANETs can combine both cellular network and WLAN to form the networks so that a WLAN is used where an access point is available and a 3G connection otherwise.

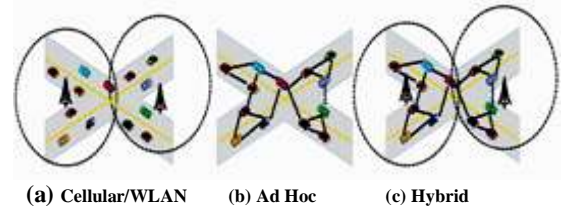

**(a)** Cellular/WLAN  **(b)** Ad Hoc  **(c)** Hybrid

**Figure 1:** Network architectures for VANETs

Stationary or fixed gateways around the sides of roads could provide connectivity to mobile nodes (vehicles), but are eventually unfeasible considering the infrastructure costs involved. In such a scenario, all vehicles and road-side wireless devices can form a pure mobile ad hoc network (Fig. 1(b)) to perform vehicle to vehicle communications and achieve certain goals, such as blind crossing.

Hybrid architecture (Fig. 1(c)) of combining infrastructure networks and ad hoc networks together has also been a possible solution for VANETs. Namboodiri et al. [9] proposed such a hybrid architecture, which uses some vehicles with both WLAN and cellular capabilities as the gateways and mobile network routers so that vehicles with only WLAN capability can communicate with them through multi-hop links to remain connected to the world. The hybrid architecture can provide better coverage, but also causes new problems, such as the seamless transition of the communication among different wireless systems.

## 3. CHARACTERISTICS OF VANET
In addition to the similarities to ad hoc networks, such as short radio transmission range, self-organization and self-management, and low bandwidth, VANETs can be distinguished from other kinds of ad hoc networks as follows[15]:

**Highly dynamic topology:** Due to high speed of movement between vehicles, the topology of VANETs is always changing.

**Frequently disconnected network:** Due to the same reason, the connectivity of the VANETs could also be changed frequently. Especially when the vehicle density is low, it has higher probability that the network is disconnected.

**Mobility modeling and predication:** Due to highly mobile node movement and dynamic topology, mobility model and predication play an important role in network protocol design for VANETs.

**Geographical type of communication:** Compared to other networks that use unicast or multicast where the communication end points are defined by ID or group ID, the VANETs often have a new type of communication that addresses geographical areas where packets need to be forwarded

**Various communications environments:** VANETs are usually operated in two typical communications environments. In highway traffic scenarios, the environment is relatively simple and straightforward (e.g., constrained one-dimensional movement), while in city conditions it becomes much more complex.

**Sufficient energy and storage:** A common characteristic of nodes in VANETs is that nodes have ample energy and computing power (including both storage and processing), since nodes are cars instead of small handheld devices.

**Hard delay constraints:** In some VANETs applications, the network does not require high data rates but has hard delay constraints.

**Interaction with on-board sensors:** It is assumed that the nodes are equipped with on-board sensors to provide information that can be used to form communication links and for routing purposes.

## 4. MOBILE AGENT

The term "mobile agent" was introduced by Telescript, which supported mobility at the programming language level. A mobile agent is a program, which represents a user in a computer network, and is capable of migrating autonomously from node to node, to perform some computation on behalf of the user. Mobile agents are defined as objects that have behaviour, state, and location [4]. Its tasks are determined by the agent application, and can range from online shopping to real-time device control to distributed scientific computing

Mobile agents are considered as a program, though they have some special properties that distinguish them from the standard programs: mandatory and orthogonal (optional) properties [5-7]. The mandatory properties are as follows:

- **Autonomy:** agents operate without the direct intervention of humans or others, and have some kind of control over their actions and internal state.

- **Decision making:** reactive or proactive decision making. Reactive decision making is implemented in some form of direct mapping from sensors input (sensing the environment) to actions using some rules. Proactive decision making may use Belief Desire Intention (BDI) architecture [6]. The beliefs of an agent are its model of the domain, its desires provide some sort of ordering between states, and its intentions are things it has decided to do.

- **Temporal continuity:** agents are continuously running processes (either running active in the foreground or sleeping/passive in the background);

- **Goal oriented:** an agent is capable of handling a task to meet its desired goal;

The orthogonal properties are as follows:

- **Mobility:** agents are capable of roaming around in an electronic network;

- **Communicative:** agents interact with other agents and (possibly) humans via some kind of agent communication language.

- **Mutual:** an agent should be capable of computing the desired tasks of the users/process by cooperating with other agents, sometimes it may reject execution of certain tasks, because (for instance) they would put an objectionable high load on the network resources or because it would cause damage to other users;

- **Learning:** agents can learn the environment factors, user preferences, etc. and develop certain degree of reasoning to take intelligent decisions/actions that improves the efficiency of the system.

### 4.1 Architectural overview & its components

Mobile agent provides a new design model for applications as compared to the traditional client server model. First and foremost, the mobile agent blows apart the very notion of client and server. With mobile agents, the flow of control actually moves across the network, instead of using the request/response architecture of client-server.In effect, every node is a server in the agent network, and the agent moves to the location where it may find the services it needs to run at each point in its execution [8]. The various components of the architecture are: -

- Agent Manager: The Agent Manager provides the communication infrastructure using the TCP/IP stack for agent transmission. It abstracts the network interface in order that agent programmers need not know any network specifics nor need to program any network interfaces.

- Security Manager: The Security Manager is responsible for identifying users, authenticating their agents, protecting server resources and ensuring the security and integrity of agents.

- Persistence Manager: The Persistence Manager is completely transparent and maintains the state of agents in transit around the network. As a side benefit, it allows for the checkpoint and restart of agents in the event of system failure.

- Event Manager: The Event Manager handles the registration, posting and notification of events to and from agents. The Event Manager can pass event notification to agents on any node in the network thus supporting agent collaboration.

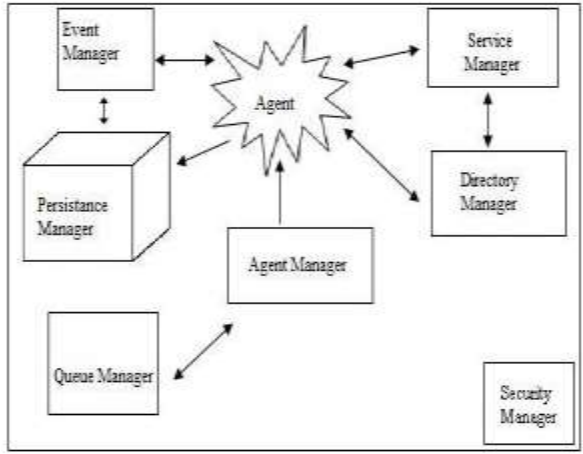

Fig 1: Architecture of a mobile agent

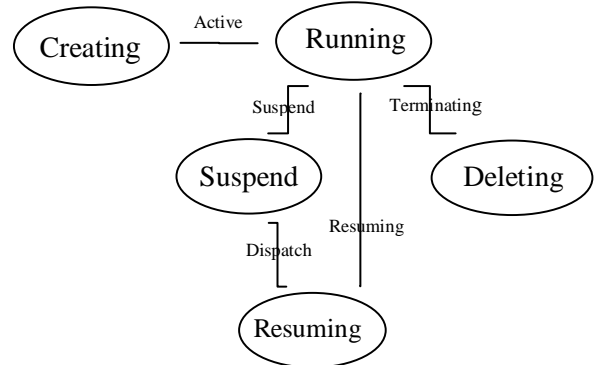

Fig 2: Life cycle of a mobile agent

- Queue Manage: The Queue Manager is responsible for the scheduling and possibly retrying the movement of agents between systems which include maintenance of agent and persistence of agent state.

- Directory Manager: The Directory manager provides naming service in the network. The Directory Manager may consult a local name service or may be set up to pass requests to other, existing name servers.

- Service manager: The Service manager provides the interface from agents to the services available at the various machines in the network. It comprises a set of programming extensions to provide access the native API's and interfacing.

### 4.2 Life Cycle of a Mobile Agent & its life state log structure

The mobile agent paradigm is based on the migrating workflow system model [9][10]. Workflow oriented life cycle model consists of five states, (creating, running, deleting, suspending, resuming) and a number of transition states. The workflow oriented life cycle model is shown in the fig. 2

To accomplish its task, the mobile agent can transport itself to another server in search of the needed resource/service, spawn new agents, or interact with other stationary agents.

Upon completion, the mobile agent delivers the results to the sending client or to another server.

1. In the creating state, the agent is created but not activated yet.
2. In the running state, the agent is running, performing actions and solve it purpose.
3. In the deleting state, the agent is terminated.
4. In the suspending state, the agent cannot run and still within the agent server.
5. In the resuming state, the agent is travelling between two server instances.

The life cycle of mobile agent begins at the moment when it is created. When PA is migrating from one host to another host in order to achieving its goals; and the PA returns its server on which it was created. Two or more than two states, in life cycle of PA, may be occurred at the different time or place. The mobile agent life state log structure can be defined in four tuple:

### 4.3 Implementation strategies

Agent mobility and naming are the two major issues in implementing the mobile agents in VANET. An agent should be uniquely named, so that its owner can communicate with or control it while it travels on its itinerary [11]. We have identified four implementation strategies:

- Sequential **CS**

This is based on the traditional client-server paradigm. The client makes a request to the first server and after processing the reply, makes a request to the second server and so on, till the list of servers to be visited is exhausted. This strategy is illustrated in figure 3 (a).

- Sequential **MA**

In this case a single MA moves from its source of origin (client) to the first site (server) in its itinerary. It then moves to the next site and so on, till it has visited all the sites in its itinerary. This strategy is illustrated in figure 3 (b).

- Parallel **CS**

This also based on the client-server paradigm. However, instead of sequential requests, the client initiates parallel threads of execution where each thread concurrently makes a request to one of the servers and processes the reply. This strategy is illustrated in figure 3 (c).

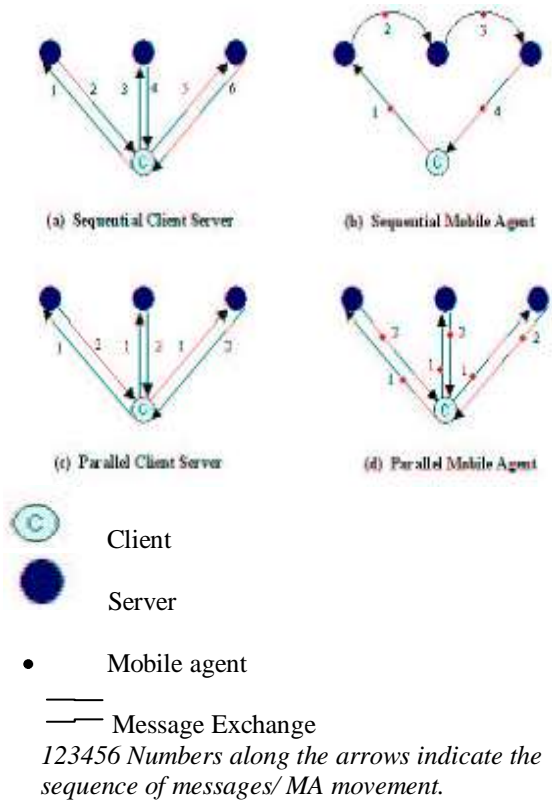

Fig. 3 Implementation strategies

- Client
- Server
- Mobile agent
- — Message Exchange

*123456 Numbers along the arrows indicate the sequence of messages/ MA movement.*

- **Parallel MA**

In this case the client initiates multiple MAs, each of which visits a subset of the servers in the itinerary. The MAs then return to the client and collate their results to complete the task. This strategy is illustrated in figure 3 (d).

## 5. MANAGING QoS USING MOBILE AGENT

Quality of service (QoS) is the measure of a service offered by the network to the user. The more deterministic network performance could be achieved with the mobile agents, so that information conceded by the network can be better delivered and network resources can be effectively utilized. A network or a service provider can put forward different kinds of services to the users. A service can be characterized by a set of requirements such as minimum bandwidth, maximum delay, maximum delay variance (jitter), and maximum packet loss rate. Quality of Service (QoS) management parameters are vital for facilitating multimedia services in a network. A typical QoS architecture should support the following: configuration, prediction, and management of QoS at all the levels of abstraction (user, system and network level); management, control, and processing of a flow must be distinct activities; application must be transparent from establishment and management; asynchronous resource management of different components; and performance enhancement. An agent based QoS architecture do supports all these features. [12]

QoS offers flexibility, scalability, efficiency, adaptability, software reusability and maintainability.

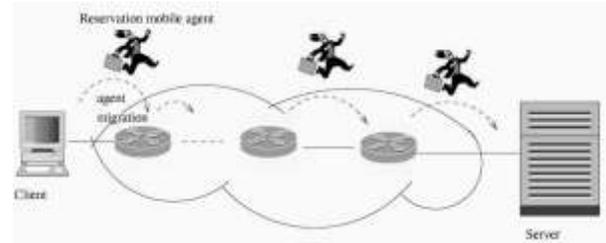

Fig 4: Agent based QoS management.

Agent-based schemes has several advantages as compared to traditional approaches: reduced latency, works in heterogeneous networks, reduced network traffic, encapsulates protocols, flexibility, adaptability, software reusability and maintainability, and facilitates the creation of customised dynamic software architectures [1,2,3]. However, mobile agent technology is still in its infancy and has certain problems that have to be resolved. Even though it is difficult to quantify these features, we explain how QoS is achieved through mobile agents.

- Flexibility: The agents allow learning capabilities to be integrated in a natural way to support delay predictions, bandwidth predictions, and play out decision-making based on the host architecture and network loads.

- Reusability: Mobile Agent software can be reused by different types of multimedia applications by making slight modifications to the software. It is possible because of autonomous operation of the agents in agent-based systems.

- Maintainability: The software can be easily maintained since every agent of an agent-based system is developed on a modular approach.

- Adaptability: Mobile agent can easily adapt the rapid changes in the network conditions (congestion/failure) and user requirements.

- Efficiency: The use of a mobile agent increases network resource utilisation efficiency because of its adaptability to network and user requirements and exchange of minimal information during task execution and decision making with multiple resource information.

- Scalability: The scalability can be achieved by using the mobile agent at a client to execute similar tasks of several users. The scheme uses only certain degree neighbour's information to compute the multiple paths.

As according to literature, a lot of concepts have already been applied on VANET to provide better QoS to the user but still some of the issues are not yet answered and they are: -

- Lack of real-time traffic information.
- Lack of access to travel information and 24 hour real-time alternate route information.
- Better alternate route guidance.

- Lack of readily available transit information to increase ridership.

Mobile agents can better address these issues in VANET. In this technique, we propose an intelligent agent based network that can provide five different types of messages in the form of agents: -

1. Direction-Finding Agent (DFA)
2. Portable Agent (PA)
3. In Vehicular Agent (IVA)
4. Observant Agent (OBA)
5. Information Finding Agent (IFA).

All vehicles moving in the network considered to be part of the network for traffic monitoring and control. The core of the system is creation of mobile agent of intelligent nodes. Agents are static nodes placed at intersections and along the streets, which maintains the database of traffic information and routing information. Road side sensors monitor the traffic situation and it is provided to nearby Agents.

The network consists of mobile nodes in an ad-hoc environment. The Portable Agents (PA) are strategically placed such that PA is well connected to at least one neighboring PA. Desirable characteristics of PA node are:

- Processing power to sustain distribution of database and node management.
- More than average normalized link capacity
- PA to PA and PA to mobile node connectivity and reliability
- Large buffer capacity to maintain ubiquitous database and routing table.

DFAs are responsible for route discovery, route maintenance and distribution of omnipresent database. Portable Agent keeps the route information and presence information. Each DFA finds the path to neighboring DFAs using PAs. PAs collect network behavior information which contain network resources like bandwidth and buffer availability at a node and pass it to a DFA. The next task is to form routing table at DFA node. To form routing table, paths from a DFA node to its neighbors are established. Path discovery is carried out by PAs. A DFA node sends Forward Portable Agents (FPAs) in network to discover the paths between itself and neighboring DFAs. Since any DFA node can be connected to nearest DFA node within maximal number of hops. This avoids flooding. Between each pair of unique DFA node, many paths are discovered and all paths are recorded in routing table. Algorithm 1 explains working of APA.

**Algorithm 1.** Advance Portable Agent (APA)

**If** (hop count== 1) **then,**

Mobile agent is deleted

**else**

if DFA node reached to next node **then**

Convey all the collected information to DFA, i.e., information regarding path followed and resources available on that path.

Create Reverse Mobile Agent with path information.

**else**

Decrease the hop count of mobile agent

Collect the network information needed for routing & submerge the mobile agents to neighbor nodes.

**end if**

Working of RPA is explained in Algorithm 2.

**Algorithm 2.** Reverse Portable Agent (RMA)

**if** DFA node reached to next node **then**

Convey all the collected information to DFA, i.e., information regarding path followed and resources available on that path to update routing table.

Remove mobile agent.

**else**

Give all the information collected to node.

Travel to next hop.

**end if**

*In Vehicular Agent (IVA)*: IVA is static agent resides in vehicle which communicates with the DFA to acquire/ spread the relevant information. IVA collects the status (moving or stationary) and location information of vehicle from sensors equipped in a vehicle.

*Observant Agent (OA):* OA is a mobile agent that travels around the network by creating its clones to propagate the decisive information during the critical situations. Examples of critical situation are accident, traffic jam, bad weather conditions, tracing a vehicle involved in crime or traffic rule violation etc. It also informs IVA and updates the vehicle database. OA is sent by DFAs to the vehicles moving in the network.

*Information Finding Agent (IFA):* IFA travels in the network to search for the requisite information as desired by vehicle user. IFA is sent by the DFA in the network on the request issued by user or DFA itself to get traffic information. [11]

## 7. Conclusion and Future Scope

Mobile agent technology has been highlighted as a very interesting approach to build applications for mobile environments. However, it is hard to find practical applications with real prototypes and using the available mobile agent platforms. One reason is probably that such platforms have been developed with a fixed distributed environment in mind, and not considering the features that may be of special interest in a mobile environment (e.g., reliance against security threats, adaptation to the network technology, and service/node discovery). Mobile agent scales effectively as the size of the data to be obtained increases.

We hope that future research and development efforts will eventually lead to consolidate a good relationship between mobile agents and mobile devices and it will also concentrate on the real time aspects.